\documentclass[12pt,a4paper]{article}

\usepackage[utf8]{inputenc}
\usepackage[T1]{fontenc}
\usepackage{lmodern}
\usepackage[english]{babel}
\usepackage[margin=1in]{geometry}
\usepackage{microtype}
\usepackage{setspace}
\usepackage{amsmath,amssymb}
\usepackage{bm}
\usepackage{graphicx}
\usepackage{booktabs}
\usepackage[round,authoryear]{natbib}
\usepackage[colorlinks=true,citecolor=blue,linkcolor=blue,urlcolor=blue]{hyperref}
\usepackage{caption}
\usepackage{xcolor}      
\usepackage{mdframed}    

\title{The atomic structure of work: a micro-action instrument reveals two-pole AI occupational exposure and its decade-scale polar inversion}

\author{%
  Shuyao Gao$^{1}$ \and Minghao Huang$^{1,\ast}$}
\date{%
  \small $^{1}$aSSIST University, Seoul, South Korea.\\
  \small $^{\ast}$Corresponding author: mhuang@assist.ac.kr}

\begin{document}

\maketitle

\begin{abstract}
\noindent
Research on artificial intelligence and work assigns each occupation a single exposure score. We build an instrument to see what those scores average over: a decomposition of 1,961 O*NET work activities into 15,817 atomic micro-actions by a consensus multi-agent LLM pipeline, clustered from text alone into seven semantic classes. Projecting exposure indicators onto these classes reveals two extreme poles, tool-mediated physical execution and planning-and-design, separated by a gap far larger than random partitions of the same data produce (permutation $P < 10^{-4}$; Cliff's $\delta = 0.80$ under our tech-risk index and $0.90$ under GPT-4 task ratings). The poles flank a broad central band that carries most work and is only weakly more compressed than chance. The poles are stable across clustering resolution, sentence encoder (under a common partition), and indicator, yet which pole is most exposed has inverted since 2013: the two extremes swap identity between the Frey--Osborne computerisation era and the LLM era, and at the occupation level an occupation's 2013 automatability declines as its linguistic content rises ($\rho = -0.40$, $n = 618$). We release the instrument and its outputs. The durable object for forecasting is the structure of work itself, not any era's exposure ranking.
\end{abstract}

\begin{mdframed}[linewidth=0.5pt,roundcorner=4pt,backgroundcolor=black!3]
\noindent\textbf{Significance Statement}\\[2pt]
\noindent
Forecasts of how artificial intelligence will affect jobs assign each occupation a single exposure number. We show that this number hides the structure that matters. Decomposing 1,961 work activities into 15,817 elementary steps and clustering them by content, we find that AI exposure is not spread along a smooth continuum but concentrated at two opposite poles, hands-on physical execution and planning-and-design, with most work in a weakly differentiated middle. Which pole is most exposed has flipped since 2013, a reversal that partly reflects what each era's tools count as automation. Exposure rankings are properties of a technological moment, not of the work itself. We release the measurement instrument for reuse.
\end{mdframed}

\section*{Introduction}

Consider a plumber dispatched to repair a burst pipe. The job, as recorded in occupational statistics, is one work activity. As performed, it is a sequence of a dozen distinct micro-actions: receive the request, extract the address and symptoms, select the right tools, travel to the site, reassure an anxious customer, locate the leak, cut the damaged section, fit and seal the replacement, test the repair, collect payment, file the record. These steps differ categorically in the kind of intelligence they demand. A large language model can already draft the scheduling message, handle the customer conversation, and issue the invoice; no deployed system can crawl under the sink. Any single number assigned to ``plumber'' averages over this heterogeneity, and it is precisely in such averages that structure disappears.

Empirical research on the labour-market impact of artificial intelligence has nonetheless converged, since \citet{frey2017future}, on the single number. An aggregate substitutability score is computed as a linear or near-linear combination of capability ratings and assigned to each occupation as a real value on $[0,1]$: a Gaussian-process classifier trained on 70 hand-labelled occupations over nine O*NET bottleneck variables \citep{frey2017future}, ability-by-application exposure sums \citep{felten2021occupational}, patent-text matching \citep{webb2020impact}, GPT-4 task ratings \citep{eloundou2023gpts}, and their cross-country \citep{gmyrek2023ilo} and agentic-AI \citep{gupta2026agentic} extensions. The shared premise is that substitutability lives on a continuum: occupations are smoothly distributed along a real-valued exposure axis, and the interpretive task is to locate each occupation on it. A parallel tradition has always worked with discrete categories instead: routine-biased technological change and the job-polarisation literature built on the routine versus non-routine dichotomy \citep{autor2003skill, autor2013polarization, goos2014explaining}, whose ``polarisation'' names hollowed wage-and-employment middles rather than the exposure geometry tested here, and Frey--Osborne's own occupation-level distribution is famously bimodal. What neither tradition has done, to our knowledge, is formally test the shape of the exposure distribution at sub-occupational resolution, even though AI capability itself is known to be jagged across tasks of similar apparent difficulty \citep{dellacqua2023jagged}. The consequence is not academic. If exposure is concentrated at two poles rather than spread along a continuum, occupations near the same aggregate score can face categorically different regimes, and every policy reading of ``moderately exposed'' misstates the structural position of the middle of the scale.

Substitution does not happen at the occupational aggregate; it happens action by action. A radiologist's ``read a chest X-ray'' may be highly substitutable while ``communicate findings to the surgical team'' is not, though both belong to the same Detailed Work Activity (DWA) and the same occupation. The task-based framework treats tasks as the smallest unit of analysis \citep{acemoglu2022task, acemoglurestrepo2018race}, and within-occupation heterogeneity is well documented: occupation-level automatability estimates collapse when task variation is admitted \citep{arntz2016risk, arntz2017revisiting}, worker self-reports show tasks varying widely within job titles \citep{autor2013tasks}, rubric-based suitability ratings find no occupation fully automatable \citep{brynjolfsson2017what, brynjolfsson2018machine}, and survey indices extend the point to generative AI \citep{henseke2025gaisi}; a PNAS synthesis has called for understanding AI's labour impact at finer resolution than the occupation \citep{frank2019toward}. Usage-data approaches map millions of AI conversations onto tasks and work activities \citep{tomlinson2025working, handa2025anthropic} or ground task labels in retrieved evidence \citep{mouchel2026evidence}. All of this work stops at or above the task. A century of work-measurement science went further long ago, decomposing jobs into elemental motions, hierarchical subtasks, and functional job elements \citep{gilbreth1917applied, annett1967task, card1983psychology, fine1971functional}, but that tradition never met AI-exposure measurement, and no study of AI substitutability has decomposed what a work activity is made of at ontology scale. A DWA-level index built from a bottleneck (Leontief-style) aggregation of technology and risk factors, introduced in our companion preprint \citep{gao2026boundedrisk}, found exposure values concentrated on the discrete set $\{0, 0.3, 0.5, 0.7, 1.0\}$ rather than smoothly filling $[0,1]$. The question this leaves open is what shape the exposure distribution takes when nothing is aggregated away.

\begin{figure}[p]
  \centering
  \includegraphics[width=\linewidth]{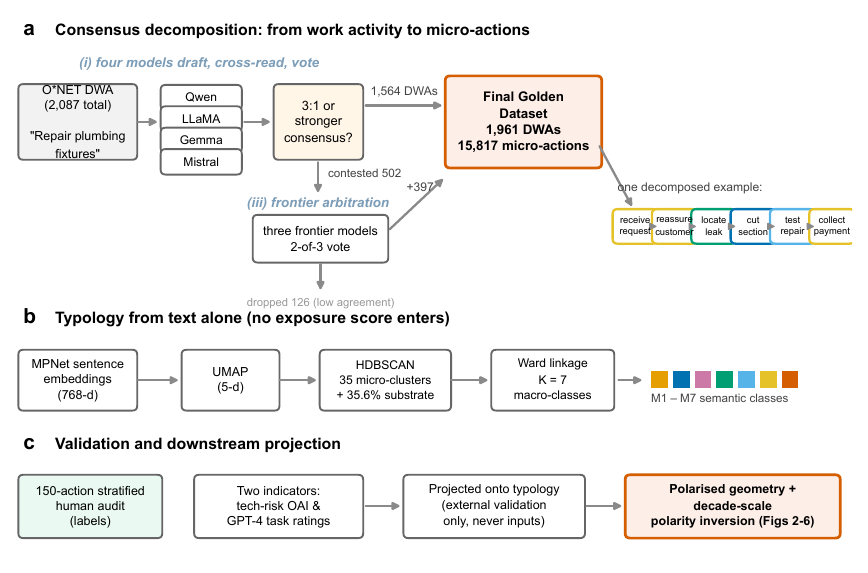}
  \caption{\textbf{From work activities to the atomic structure of work.} \textbf{a}, Consensus decomposition: each O*NET Detailed Work Activity (DWA) is independently drafted into a candidate micro-action sequence by four open-weight LLMs, which then cross-read and vote. Consensus of 3:1 or stronger accepts 1,564 DWAs directly; the 502 contested cases are arbitrated by three frontier-class models (2-of-3), adding 397; 126 low-agreement cases are discarded. The result is 15,817 micro-actions across 1,961 DWAs; the strip shows one decomposed example, coloured by eventual macro-class. \textbf{b}, The typology is built from action text alone: MPNet sentence embeddings, UMAP reduction, HDBSCAN micro-clustering (35 micro-clusters plus a 35.6\% generic substrate), and Ward linkage into seven macro-classes. No exposure score enters the construction. \textbf{c}, Validation and downstream projection: a 150-action stratified human audit validates the downstream intelligence-type labels (Methods); the decomposition itself is quality-controlled by inter-model consensus. Exposure indicators built by different procedures are projected onto the typology as external validation only.}
  \label{fig:schematic}
\end{figure}

Here we introduce the \textbf{micro-action} as a unit of analysis for AI exposure.\footnote{Throughout, a micro-action is the smallest purposeful step of occupational work. It is distinct from ``micro-action recognition'' in video understanding, where the term denotes subtle, involuntary body movements. We use \emph{occupational micro-actions} where disambiguation matters.} We decompose the 1,961 DWAs of the O*NET 30.2 ontology into 15,817 atomic micro-actions using a consensus multi-agent LLM pipeline in which four open-weight models independently draft, cross-read, and vote, frontier-class models arbitrate contested cases, and low-agreement cases are discarded (Fig.~\ref{fig:schematic}; Methods); an independent blind human audit confirms the downstream typology is reproducible (Fleiss $\kappa = 0.90$) while flagging the automatic per-action labels as the noisier layer (Methods). Sentence embeddings of every micro-action are reduced with UMAP and clustered, first into 35 micro-clusters by HDBSCAN and then into seven interpretable macro-classes (M1--M7) by hierarchical Ward linkage. The typology is built from text alone: no exposure score of any kind enters its construction.

We then project two exposure indicators built by different procedures onto this typology: the bounded tech-risk Automation Index (OAI) of our companion preprint \citep{gao2026boundedrisk}, and the GPT-4 task ratings of \citet{eloundou2023gpts}. Both reveal the same \textbf{two-pole structure}. Tool-Mediated Physical Execution (M2, mean OAI $0.054$) and Planning \& Design (M7, mean OAI $0.499$) form two extremes separated by Cohen's $d = 2.41$; the six middle classes form a low-contrast band, not a smooth gradient. The geometry survives four independent stress tests: a permutation null (the polar gap exceeds all of 20,000 random partitions of the same data), clustering resolution (the polar gap widens from $0.45$ at $K=7$ to $0.57$ at $K=15$), sentence-embedding encoder (under a common $k$-means partition the polar gap is $0.47$ for both MPNet and BGE embeddings; the density-based pipeline that yields the substrate is MPNet-specific, Methods), and indicator choice (Cliff's $\delta = 0.902$ for the extreme pair under the GPT-4 ratings). Against this spatial stability sits a temporal inversion: the two exposure extremes swap identity between Frey--Osborne's 2013 computerisation era and the LLM era, and at the occupation level 2013 automatability falls as linguistic content rises ($\rho = -0.40$, $n = 618$). It is the extremes that invert; the overall ranking decouples rather than mirroring. A four-way intelligence-type classification of the actions (Linguistic, Multimodal Perception, Embodied, and Human-Bound) supplies the mechanism: the era's dominant AI capability determines which intelligence type maps to the high-exposure pole, so any single-point exposure forecast inherits the era's capability frontier and reverses with it.

The contribution is a reframing of what the empirical object of AI-and-work research should be. The continuous gradient that has organised the field is an artifact of aggregation: summing many capability ratings produces smooth output even when the substrate has two extreme poles. What is durable, across clustering resolutions, indicators, and a decade of capability change, is the geometry: two poles, a low-contrast middle, and roughly one third of all work consisting of a generic action substrate present in every occupation (mean per-occupation share 0.34, interquartile range 0.25--0.44). The two-pole gap additionally survives a sentence-encoder swap; the generic substrate, which only the density-based pipeline defines, is established under a single encoder (Supplementary Note~1). That geometry, not any era's exposure ranking, is the structure a forecast can safely stand on.

\section*{Results}

\subsection*{An atomic typology of work}

\begin{figure}[p]
  \centering
  \includegraphics[width=0.95\linewidth]{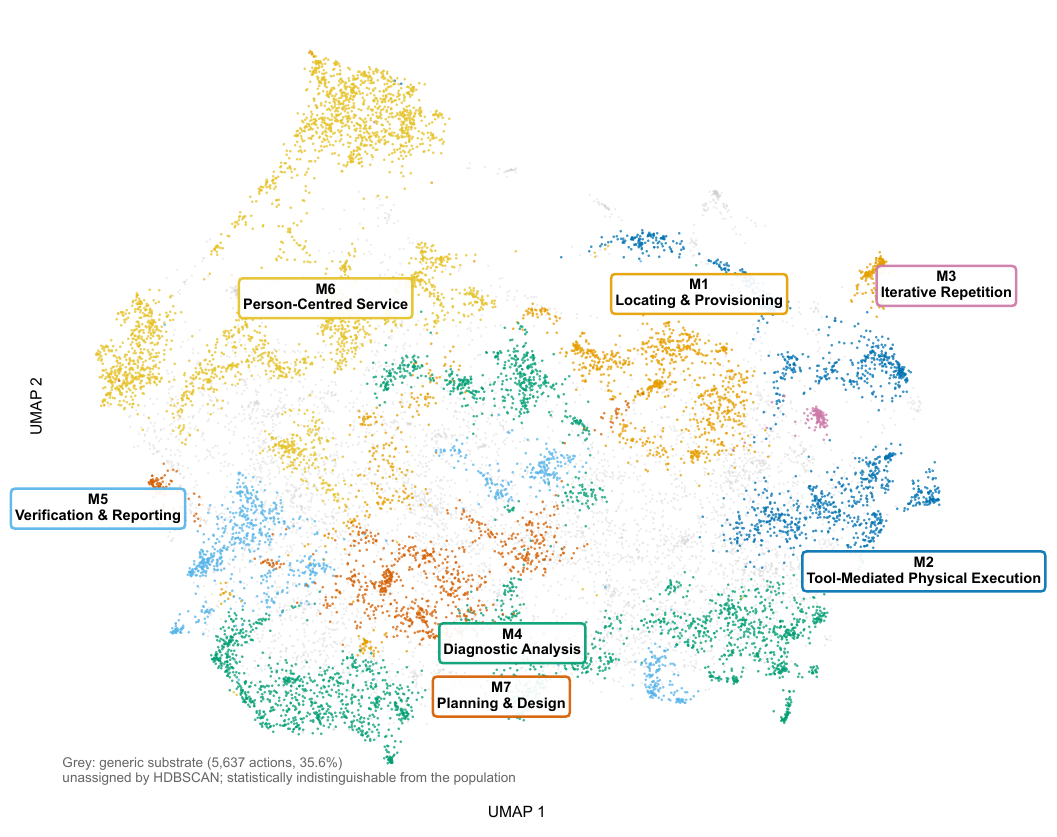}
  \caption{\textbf{The atomic map of work.} UMAP projection of all 15,817 micro-actions, coloured by macro-class assignment ($K=7$ Ward cut over 35 HDBSCAN micro-clusters); labels mark class medians. The generic-action substrate (light grey; 5,637 actions, 35.6\%) is plotted beneath the class layers. The typology is constructed from action text alone; no exposure indicator enters its construction.}
  \label{fig:umap}
\end{figure}

The decomposition pipeline (Methods) yields a Final Golden Dataset of 1,961 DWAs decomposed into 15,817 micro-actions (mean 8.07 actions per DWA; range 5--13), each carrying a free-text description, a canonical process stage (Intent Communication, Navigation/Addressing, Perception/Diagnosis, Manipulation/Execution, or Feedback/Verification), a cognitive-versus-physical tag, and its ordinal position in the sequence. Clustering the action descriptions with MPNet sentence embeddings, UMAP reduction, HDBSCAN micro-clustering, and Ward macro-clustering produces seven macro-classes plus two structural groups (Fig.~\ref{fig:umap}; Supplementary Table~1). The seven classes are M1 Locating \& Provisioning (8.5\% of actions), M2 Tool-Mediated Physical Execution (8.6\%), M3 Iterative Repetition (0.7\%, a meta-action layer of ``repeat steps'' constructs), M4 Diagnostic Analysis (16.3\%), M5 Verification \& Stakeholder Reporting (6.2\%), M6 Person-Centred Service Interaction (18.4\%, the largest), and M7 Planning \& Design (5.6\%). The two structural groups are a generic-action substrate of 5,637 actions (35.6\%) that HDBSCAN declines to assign to any cluster, and a small mixed-DWA cell (96 DWAs, 4.9\%) whose action mass is spread too evenly for any class to dominate.

The seven classes differentiate simultaneously along three descriptive axes (Supplementary Fig.~1). Each of the five canonical process stages is the dominant stage of at least one class; the cognitive share of actions spans from 28.4\% (M2) to 84.2\% (M7); and mean sequence position spans from 3.97 (M7, concentrated at workflow onsets) to 6.99 (M3, concentrated at tails). No two classes occupy the same cell of this cross-product (the stage, cognitive-physical, and step-order shares are themselves inputs to the macro-clustering stage, so their differentiation is partly by construction; re-clustering on the semantic centroids alone reproduces significant cognitive-physical and stage-dominance differentiation, but not sequence-position differentiation; Supplementary Note~1). The plumber's sequence from the opening spans this map: reassuring the customer is the territory of M6, locating the leak of M4, cutting and fitting of M2, testing of M5, while the scheduling and payment steps are the kind of generic actions the substrate collects.

\subsection*{Two exposure poles, far beyond chance}

\begin{figure}[p]
  \centering
  \includegraphics[width=\linewidth]{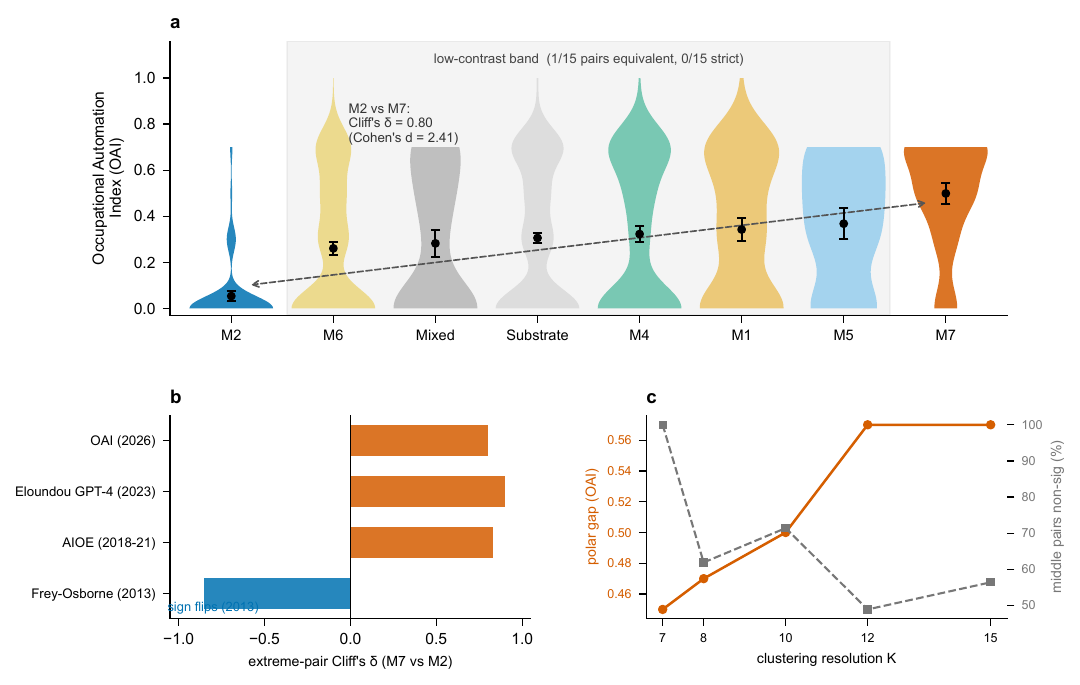}
  \caption{\textbf{Two exposure poles flanking a central band.} \textbf{a}, DWA-level OAI by analysis group (violins; black points, mean $\pm$ 95\% CI), ordered by mean. M2 (Tool-Mediated Physical Execution) and M7 (Planning \& Design) anchor the extremes (Cohen's $d = 2.41$, Cliff's $\delta = 0.799$). Within the shaded middle band, all 15 pairwise contrasts fail Bonferroni-corrected significance while only 1 of 15 passes equivalence testing at $d = 0.2$ (0 of 15 at $\pm 0.05$). \textbf{b}, The extreme-pair contrast (Cliff's $\delta$, M7 vs M2) under four independent exposure indicators; the sign flips under Frey--Osborne (2013). \textbf{c}, Resolution sweep: re-cutting the dendrogram at $K = 7$--$15$ widens the polar gap from 0.45 to 0.57 (orange) while middle pairs gradually resolve (grey).}
  \label{fig:polarised}
\end{figure}

We projected the DWA-level OAI onto the eight analysis groups (M3 has no DWA in which it dominates and is excluded; Methods). The eight OAI distributions are not interchangeable (Kruskal--Wallis $H = 172.88$, $p = 6.2 \times 10^{-34}$). Ordered by mean, the groups are anchored by two extremes: M2 at mean OAI $0.054$ ($n = 170$ DWAs) and M7 at $0.499$ ($n = 99$), separated by Cohen's $d = 2.41$ and Cliff's $\delta = 0.799$ (Bonferroni-corrected Mann--Whitney $p = 1.1 \times 10^{-33}$). The six remaining groups cluster within a band $0.107$ wide, one tenth of the scale (Fig.~\ref{fig:polarised}).

Ordering the means of any bounded variable places two of them at the extremes by construction, so we tested the poles against exactly that null. Randomly re-partitioning the 1,961 DWAs into eight groups of the observed sizes, the polar gap (largest minus smallest group mean) averages $0.074$ and never reaches the observed $0.445$ in 20,000 permutations ($P < 5\times10^{-5}$); the same holds under the independent GPT-4 ratings ($P < 5\times10^{-5}$). The two poles are therefore a property the text-based clustering recovers, not an artifact of ordering group means. We are deliberately narrower about the middle: the central band is at most weakly more compressed than random partitions ($P = 0.04$ under OAI, $0.14$ under GPT-4), and the eight group means are not more bimodally distributed than chance ($P = 0.84$). The finding is two extreme poles flanking a broad central mass of work, not a bimodal distribution of exposure.

The pairwise structure is stark: of 28 Bonferroni-corrected pairwise contrasts, exactly 13 are significant, and every one of them involves M2 or M7. All 15 pairs among the six middle groups fail to reject the null. Failure to reject is not evidence of equivalence, so we ran two-one-sided equivalence tests (TOST) on the middle pairs at the conventional smallest-effect margin of Cohen's $d = 0.2$: only 1 of 15 pairs reaches equivalence, and 0 of 15 at a stricter absolute margin of $\pm 0.05$. The middle is therefore a \textbf{low-contrast band}, a gentle gradient that neither the significance tests nor the equivalence tests can flatten. It is neither a staircase nor a plain. The two continuous indicators do resolve more of the middle pairs (Supplementary Table 1), so instrument resolution contributes to the band's breadth alongside any property of the work itself.

The two-pole shape is not a property of the $K=7$ resolution. Re-cutting the dendrogram at $K = 8, 10, 12, 15$ widens the polar gap monotonically (from $0.45$ to $0.57$) while the middle band gradually resolves into distinguishable sub-clusters (Fig.~\ref{fig:polarised}c). The sweep also exposes the sharpest internal critique of our own headline cut: at $K=12$, the Diagnostic Analysis class M4 splits into three sub-classes whose mean OAI spans from $0.031$ (physical equipment monitoring) to $0.604$ (data analysis), an internal range wider than the headline polar gap itself. Lexical similarity (``inspect'', ``examine'', ``review'') held together actions that differ on every dimension that matters economically. The \emph{chimera} is a check on our own cut: text-similarity neighbourhoods must be validated against substantive moments computed outside the embedding space (Supplementary Note~2). The two-pole shape itself survives the split, because the sub-classes simply take over the polar roles. The stable object is the two-poles-plus-band topology, not the identity of any particular class.

\subsection*{The shape reproduces under independent indicators}

The two-pole shape could in principle be an artifact of the OAI's construction, but three exposure indicators computed entirely outside our pipeline reproduce it on the same typology (Supplementary Fig.~S8). The GPT-4 task ratings of \citet{eloundou2023gpts}, projected to DWAs through the O*NET task--DWA crosswalk, yield $H = 298.4$ and an extreme-pair Cliff's $\delta = 0.902$, larger than under our own OAI, with 12 of 15 middle pairs non-significant. At the DWA level the two indicators, built by wholly different procedures, agree at Spearman $\rho = 0.635$ ($n = 1{,}958$); both, however, are LLM-rated, so the two fully non-LLM instruments below provide the stronger independence check. Felten's AIOE \citep{felten2021occupational} reproduces the two-poles-plus-band form ($H = 287.4$), as does Frey--Osborne's computerisation probability \citep{frey2017future} ($H = 329.0$). Under Frey--Osborne, however, the poles are \emph{swapped}, a sign reversal we return to below. The shape lives in the action-level typology, not in any rating instrument. One limit is structural, and it is inseparable from the novelty. Resolving work below the task is only newly possible because language models can decompose it; no prior instrument, LLM-based or not, operates at the action level across an entire occupational ontology (manual time-and-motion studies decomposed single jobs a century ago, but none scaled to a full economy), so there is as yet no non-LLM action-resolved exposure indicator to triangulate against. The two action-resolved indicators (our OAI and Eloundou's ratings) are LLM-rated, while the two non-LLM indicators (AIOE and Frey--Osborne) are occupation-level scores replicated across each occupation's DWAs. The action-level geometry is therefore established under LLM-rated action indicators and corroborated, at the occupation and class level, by non-LLM ones. The decisive independent check, expert or worker ratings of the micro-actions themselves, was not even definable before an action-level decomposition existed; the instrument we release makes it possible for the first time.

\subsection*{One third of work is a generic substrate}

The 5,637 actions (35.6\%) that HDBSCAN leaves unclustered are not noise in the pejorative sense; they are a structural layer. On every feature axis available to us, this group is statistically indistinguishable from the full population: its process-stage distribution matches to within 1.2 percentage points on every stage, its cognitive share differs by 1.2 points, its mean sequence position by 0.04 steps, and its OAI distribution is indistinguishable from the population's (Kolmogorov--Smirnov $D = 0.027$, $p = 0.85$). These are the domain-neutral verbs of work (review, document, ensure, follow up), present in every occupation and constitutive of none.

This substrate has a sharp economic implication. It is present in every occupation: the per-occupation substrate share averages $0.34$ (s.d.\ $0.14$, interquartile range $0.25$--$0.44$), and the aggregate share is hyperparameter-stable, staying within $0.33$--$0.41$ across an HDBSCAN settings grid (Supplementary Note~8). Occupation-level indices therefore average over a large generic component everywhere, and because the share itself varies, substrate compression would additionally bear hardest on the occupations carrying the largest generic layers. On its face this is also the layer current language models can most readily perform end to end. This is a hypothesis, not an indicator-backed finding: the OAI rates the substrate only at the population mean ($0.300$, statistically indistinguishable from the whole) and, being a DWA-level index, cannot score the generic verbs directly, so no measurement in hand establishes that the substrate is more performable than average. If it holds, substitution concentrated in the substrate would compress roughly one third of the action content of every occupation simultaneously (a count of decomposition steps, not of work time; micro-actions carry no duration weights). The result would be a within-occupation flattening that preserves job titles while shrinking support layers, not the cross-occupation displacement wave that exposure rankings predict. Occupation-level indices, ours included, were not built to detect this dynamic; per-occupation time-budget decomposition into substrate versus domain-specific shares is the datum the field should collect next.

\subsection*{Intelligence types explain the poles}

Why are M2 and M7 the poles, and why does the middle refuse to be ordered? We classified each of the 15,817 actions into one of four intelligence types (Linguistic, Multimodal Perception, Embodied, or Human-Bound) by prototype-anchored nearest-centroid assignment under a BGE encoder. Three annotators blind to the model labels and to the study hypotheses reproduce the four-category construct at Fleiss $\kappa = 0.90$, but the automatic labels match these annotators only at $\kappa = 0.63$ (substantial; Methods), so we read the intelligence-type layer at the class level rather than trusting any single action's label. The classification is independent of the macro structure and of every exposure indicator.

The four types partition the exposure scale far more cleanly than the semantic classes do. Assigning each DWA its modal intelligence type yields mean OAI of $0.427$ for Linguistic-dominant DWAs ($n = 1{,}070$), $0.219$ for Human-Bound ($n = 238$), $0.125$ for Multimodal Perception ($n = 331$), and $0.060$ for Embodied ($n = 322$): Kruskal--Wallis $H = 527.6$ ($p = 4.9 \times 10^{-114}$), with all six pairwise contrasts Bonferroni-significant (Supplementary Fig.~S5). The Linguistic class's mean exposure is $3.37\times$ the mean of the other three classes combined on the full DWA-modal basis; Supplementary Note~5 recomputes this lead on the 150-action blind-audit sample, where it is a smaller $2.5\times$ ($2.2\times$ after propagating encoder error), the two figures differing because they use different denominators. The result replicates under an independent encoder family (Methods).

The intelligence-type composition of the classes then explains the geometry. M7 is 95\% Linguistic; M2 is 56\% Embodied and only 15\% Linguistic (Supplementary Fig.~S5b). The poles are the classes whose composition is most concentrated in the highest- and lowest-exposure intelligence types. The middle classes are mixtures; M6, for instance, is 41\% Linguistic and 43\% Human-Bound. Mixtures are structurally what middling aggregate scores look like at the action level. The low-contrast band is low-contrast because it is compositionally blended, not because its content is homogeneous.

\subsection*{A decade-scale polarity inversion}

If polar identity is mediated by which intelligence type the era's dominant AI capability targets, and the dominant capability moves, the polarity should move with it. The thirteen-year record shows that it did. We projected four indicators spanning thirteen years of capability change (Frey--Osborne 2013, AIOE 2018--2021, GPT-4 ratings 2023, and OAI 2026) onto a nine-group partition, splitting the chimeric M4 into its three sub-classes (Methods), and compared polar orderings (Table~\ref{tab:inversion}).

M2, the \emph{highest}-exposure class under Frey--Osborne (mean probability $0.703$), is the second-\emph{lowest} under the 2026 OAI ($0.058$); M7 moves the opposite way, from $0.287$ to $0.504$. The equipment-monitoring sub-class M4-HVAC falls from $0.654$ to $0.035$ (rank 8 to rank 1 of 9) and the data-analysis sub-class M4-data rises from $0.340$ to $0.601$ (rank 4 to rank 9): the four largest movers are exactly the four poles. It is the \emph{extremes} that invert. The cleanest and most disaggregated evidence for the mechanism is at the occupation level, where across $618$ matched codes a 2013 computerisation probability declines as an occupation's Linguistic composition rises ($\rho = -0.40$, $P \approx 10^{-25}$): the work rated most automatable in 2013 was precisely the least linguistic, and Linguistic is the intelligence type the LLM era most exposes. Aggregated to class means, the Frey--Osborne ordering runs opposite to the LLM-era orderings (nine-class Spearman $\rho = -0.750$, $p = 0.020$ against Eloundou; $\rho = -0.650$, $p = 0.058$ against OAI). A within-class bootstrap yields intervals excluding zero ($[-0.85, -0.55]$ and $[-0.70, -0.33]$), but it conditions on the nine fixed classes and does not carry the uncertainty of having only nine, so we treat the permutation $p$ as the conservative figure. We read this nine-point class correlation as a compact summary of the pole swap, not as a distribution-wide mirror image: at the DWA level ($n = 1{,}958$) the two eras \emph{decouple} rather than reverse (Frey--Osborne versus OAI $\rho = -0.116$; the raw occupation-level correlation is likewise $\rho = -0.08$), while the two LLM-era indicators agree at $\rho = +0.635$. The pole swap is already present under the original eight-group partition, and splitting M4 sharpens rather than creates it. A provenance check rules out data artefacts: our Frey--Osborne values are parsed directly from the original Oxford Martin appendix and match the compilation circulating in the literature at $\rho = 1.000$ across 653 matched occupation codes.

The 2013 assessment was not carelessly made; it faithfully reflected an era whose engineering consensus placed the automation frontier at structured physical execution and its durable bottlenecks at language and unstructured planning. The subsequent decade moved the frontier to language, and the polar ordering reversed accordingly. Instruments anchored to successive capability frontiers rank-reverse; any single-point exposure forecast inherits its era's frontier.

\begin{table}[p]
\centering
\small
\caption{\textbf{Per-class mean exposure under four indicators spanning 2013--2026.} Nine-group partition (M4 split into its three sub-classes), ordered by mean OAI ascending. The four largest era-to-era movers are the four poles: M2 and M4-HVAC fall from the Frey--Osborne high pole to the LLM-era low pole; M4-data and M7 rise in the opposite direction. Macro-level Spearman between Frey--Osborne and Eloundou orderings: $\rho = -0.750$, $p = 0.020$. Class means under this nine-group partition differ slightly from the eight-group values in the text (e.g.\ M2 $0.058$ vs $0.054$) because splitting M4 reassigns the dominant class of some DWAs.}
\label{tab:inversion}
\begin{tabular}{lrrrr}
\toprule
\textbf{Class} & \textbf{OAI 2026} & \textbf{Eloundou 2023} & \textbf{AIOE} & \textbf{Frey--Osborne 2013} \\
\midrule
M4-HVAC                     & 0.035 & 0.214 & $-0.601$ & 0.654 \\
M2 (Tool-Mediated Physical) & 0.058 & 0.086 & $-0.787$ & 0.703 \\
M4-patrol                   & 0.132 & 0.357 & $0.112$  & 0.388 \\
M6 (Person-Centred Service) & 0.262 & 0.380 & $0.357$  & 0.300 \\
Generic substrate           & 0.300 & 0.340 & $-0.008$ & 0.471 \\
M1 (Locating \& Provisioning)& 0.345 & 0.387 & $-0.192$ & 0.626 \\
M5 (Verification \& Reporting)& 0.350 & 0.464 & $0.574$  & 0.306 \\
M7 (Planning \& Design)     & 0.504 & 0.497 & $0.469$  & 0.287 \\
M4-data                     & 0.601 & 0.543 & $0.890$  & 0.340 \\
\bottomrule
\end{tabular}
\end{table}

\begin{figure}[p]
  \centering
  \includegraphics[width=\linewidth]{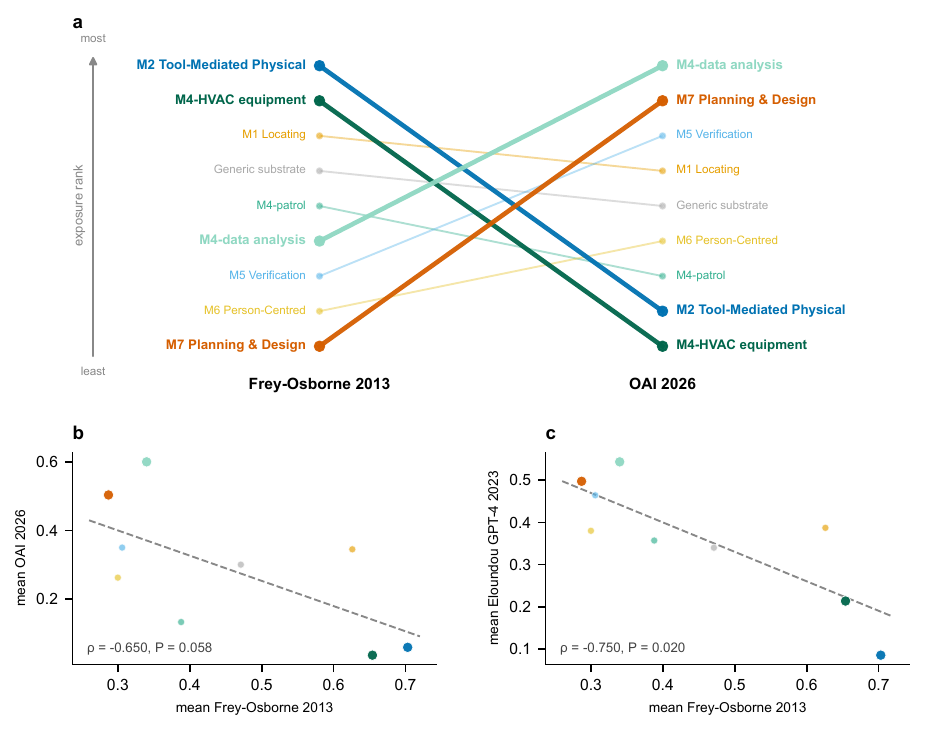}
  \caption{\textbf{The decade-scale polarity inversion.} \textbf{a}, Exposure rank of the nine classes (M4 split into its three sub-classes; Methods) under Frey--Osborne 2013 and the 2026 OAI. The four largest movers are the four poles: M2 and M4-HVAC fall from most-exposed to least-exposed while M4-data and M7 rise in the opposite direction. \textbf{b},\textbf{c}, Class-mean Frey--Osborne exposure against the 2026 OAI (\textbf{b}) and the 2023 GPT-4 task ratings (\textbf{c}); both relationships slope downward (Spearman $\rho = -0.650$ and $-0.750$).}
  \label{fig:inversion}
\end{figure}

\clearpage

\section*{Discussion}

A structural reading follows. Occupation-level exposure scores average over categorically different work: at the action level, two extreme poles (separated by $d = 2.41$ and far beyond chance) flank a broad central band, with a generic substrate running beneath every occupation. Where an index is a smooth summary of many rated dimensions, this internal structure is invisible by construction; the point is not that all aggregation linearises (our own OAI is a non-linear bottleneck aggregate that already departs from a smooth gradient at the DWA level) but that any occupation-level summary discards the action-level poles. Indices reporting ``moderate'' exposure for an occupation are reporting its location in the central band, the region where exposure measurement discriminates most weakly.

The two findings, spatial stability and temporal inversion, admit a compact formal statement in the task-content-of-production tradition \citep{acemoglu2011skills}. Let occupation $j$ be represented by its composition over the four intelligence types, $\bm{\omega}_j = (\omega_{L,j}, \omega_{M,j}, \omega_{E,j}, \omega_{H,j})$ with $\sum_k \omega_{k,j} = 1$. This vector is a property of how the occupation's work is structured, not of the technology available to substitute for it; it is the durable object our decomposition recovers, measured here at a single 2026 vintage (task content itself drifts as occupations shed and gain work \citep{autor2024new}, so $\bm{\omega}_j$ is durable relative to $\bm{\Psi}$'s rotation, not immutable). Era-specific substitutability is then a projection,
\begin{equation}
  S_j(t) \;=\; \textstyle\sum_{k} \omega_{k,j}\, \Psi_k(t)\, \Gamma_k(t),
  \label{eq:model}
\end{equation}
where $\Psi_k(t)$ is the technical capability frontier for intelligence type $k$ at time $t$ and $\Gamma_k(t)$ is the era's institutional risk tolerance for deploying that capability. Between 2013 and 2026, $\Psi_L$ and $\Psi_E$ approximately exchanged ranks: language modelling crossed its bottleneck while general-purpose manipulation did not. The era-to-era change in exposure is therefore dominated by $\omega_{L,j} - \omega_{E,j}$. M7, at $\omega_L - \omega_E = 0.92$, rose furthest; M2, at $-0.41$, fell furthest. This is precisely the observed reversal. The \emph{stable geometry} of our title is the matrix $\{\bm{\omega}_j\}$; the \emph{reversing poles} are the rotation of $\bm{\Psi}(t)$ across it. A future era in which embodied AI matures would rotate the poles again while leaving the two-pole shape intact. We present equation~\eqref{eq:model} as an accounting identity that organises this rotation, not a fitted model (its capability and risk terms enter only through their product; Supplementary Note~6). Its content is the direction it predicts, that era-to-era exposure change tracks $\omega_{L,j} - \omega_{E,j}$. Regressing the class-level exposure change on $\omega_L - \omega_E$ recovers this positive slope ($0.57$), and it persists with the four pole classes removed ($R^2 = 0.83$ across the five middle classes), so it is not a two-corner artifact. But $\Delta S$ (a language-favouring minus a physical-favouring index) and $\omega_L - \omega_E$ (a language minus physical share) are both language-versus-physical contrasts, so this confirms that the identity's arithmetic holds across classes rather than estimating an independent relationship; the substantive content is the sign and slope, not the fit (Supplementary Note~6).

The instruments also differ in scope, not only in vintage. Frey--Osborne scored computerisation broadly, including mobile robotics, while the 2023--2026 indicators score generative-AI exposure and some rate embodied tasks unexposed by construction, so part of the sign flip is built into what each instrument counts as automation. In our terms that is the point rather than an objection to it: an exposure instrument encodes its era's capability frontier, so scope choice and frontier belief are the same commitment \citep{autor2015why}. Three results establish that the inversion is more than definitional. The 2013 ranking carried real signal: over the following decade, employment in the occupations Frey--Osborne rated most exposed grew rather than collapsed, though more slowly than in low-risk occupations \citep{georgieff2021what}, so what reversed is the ranking's \emph{polar reading} rather than its content. Felten's AIOE, crowd-rated over ten cognitive AI applications in 2018--2021 with no language model in its pipeline, already places the poles in the LLM-era ordering (Supplementary Fig.~S8); it is the only pole-ordering instrument carrying no model-generated ratings, though its cognitive applications include language-related abilities and so are not fully independent of the linguistic frontier. And task-content drift, the remaining vintage confound, can be bounded directly: reprojecting the 2013 ratings through the occupation--DWA structure of O*NET 18.1 (March 2014, the earliest release carrying the detailed-work-activity layer) rather than the 2026 structure leaves the per-DWA values essentially unchanged (Spearman $0.91$, mean absolute difference $0.07$, $97.8\%$ coverage) and the M2--M7 pole swap identical (Supplementary Note~8). The scope difference itself is what remains: a strictly 2013-contemporaneous decomposition would require re-decomposing pre-2014 activities that the ontology did not yet carry \citep{autor2024new}.

This reframing has direct consequences for the forecasting practice that dominates public discussion. Capability-stage roadmaps and displacement projections \citep{morris2023levels, openai2024fivelevels, chui2023mckinsey, ellingrud2023mckinsey} share one structural property: each takes a contemporaneous capability snapshot and projects occupation-level exposure forward. The polarity inversion is the existence proof that this pattern has already failed once at a thirteen-year horizon, not through poor execution but because the pattern itself anchors on the wrong invariant. What transports across capability eras is the geometry: the number of poles, the low-contrast band, the substrate share, and the intelligence-type composition $\bm{\omega}_j$ of each class. The productive inferential direction is therefore reversed: rather than asking which occupations tomorrow's capability exposes, ask which intelligence type the frontier is moving toward, and read off the affected pole from a stable map. The policy stakes are concrete: guidance built on the 2013 ordering pointed workers toward the cognitive pole that is now the most exposed, and reskilling subsidies allocated on exposure rankings inherit exactly this fragility.

The generic substrate sharpens the point about where substitution will bite first. Because every occupation carries a substantial generic layer (per-occupation share $0.34 \pm 0.14$), current LLM capability acting on it predicts \emph{flattening}: compression of generic action content within nearly every occupation, visible as fewer support roles per substantive role, rather than the extinction of specific occupations. Early evidence of contracting entry-level employment in exposed occupations is consistent with this pattern \citep{brynjolfsson2025canaries}. A full regime map of augmentation, structural protection, and substitution across the geometry is given in the Supplementary Information. Cross-sectional exposure indices largely average over this layer, and its dispersion across occupations (s.d.\ $0.14$) is a margin they were never designed to read. This is a concrete, testable prediction that distinguishes our account from the displacement-wave tradition: white-collar adjustment under generative AI should appear in within-occupation time-use and staffing-ratio data (for example, ATUS time diaries, JOLTS staffing flows, and O*NET incumbent task-frequency updates) before it appears in occupational employment counts.

The reversal also redistributes exposure across social groups. Across the 22 major occupation groups, a group's pole position ($\omega_L - \omega_E$) rises steeply with its required educational preparation (O*NET Job Zone, Spearman $\rho = 0.83$; $\rho = 0.70$ across all 923 occupations) and with its median earnings ($\rho = 0.72$, from about \$34{,}000 at the Embodied pole to \$80{,}000 at the Linguistic pole), so the LLM era's high-exposure pole is the \emph{more}-educated and \emph{better}-paid one, inverting the historical pattern in which measured automation exposure concentrated on lower-paid manual work. Exposure does not track sex along the pole axis ($\rho = 0.23$, not significant; the most female-dominated groups sit in the central band, not at either pole). It does differ by race, and that difference decomposes. The cognitive pole is more White ($66\%$ versus $51\%$) and Asian ($9\%$ versus $5\%$) and the manual pole more Hispanic ($25\%$ versus $12\%$; ACS 2022 microdata, $\approx 2.0$ million U.S.\ workers), but holding educational preparation fixed the White- and Black-share associations vanish (partial $\rho = 0.06$ and $0.07$) while the Asian and Hispanic associations survive at roughly half strength ($0.26$ and $-0.20$). The racial gradient is thus mostly the education gradient seen through occupational segregation \citep{alonsovillar2013occupational}, not a separate channel. What this measures is the composition of exposure rather than an incidence of harm, since whether exposure becomes displacement or augmentation depends on which regime each pole occupies; the overlay characterises the poles, not the middle band or the substrate where substitution is predicted to concentrate, and the U.S.\ census categories it uses do not transfer abroad (Supplementary Note~7).

Three limitations bound these claims (Supplementary Note~8). The micro-actions are LLM-generated and were produced without worker participation: cross-model consensus and the independent human audit constrain but cannot eliminate biases shared across models trained on overlapping corpora, and components of work that resist stepwise description (emotional labour, continuous vigilance, tacit adaptation) are undersampled, so replication with expert- or survey-derived decompositions would strengthen the base. The ontology is US, English-language, and formal-economy, so the geometry should not yet be extended to economies with large informal or manufacturing sectors. And the analysis is a structural cross-section: it characterises exposure, whose translation into displacement or augmentation depends on adoption and institutions outside our data. Two narrower caveats: the class-level inversion statistic rests on nine classes, though the occupation-level ($n = 618$) and DWA-level ($n = 1{,}958$) analyses that anchor the temporal claim do not depend on that partition; and the Human-Bound type bundles interpersonal demand, whose value has risen with technology as a complement rather than a substitute \citep{deming2017growing}, with institutional accountability (capability and $\Gamma$ in equation~\eqref{eq:model}), so the released per-class confidences allow soft reassignment. Exemplar occupations for every class are in the Supplementary Information.

The unit-of-analysis argument generalises beyond AI exposure. Occupational-mobility models, skill-graph constructions, and embedding-based classification systems all measure proximity in a semantic space induced by how people \emph{describe} work; the M4 chimera demonstrates that description-space proximity and economic exposure can diverge arbitrarily. Micro-action decomposition offers these literatures what it offered exposure measurement here: a substrate fine enough that the structure of interest is no longer averaged away before the analysis begins. The atomic structure of work is measurable at scale, it is stable where exposure rankings are fragile, and it is the object on which a cumulative empirical science of AI and work can be built.

\section*{Methods}

\subsection*{Data construction}

We took as our base the O*NET 30.2 release of the US occupational ontology, which catalogues 923 detailed occupations linked to 2,087 Detailed Work Activities (DWAs). DWAs are the finest grain at which O*NET labels work; below them the ontology is silent.

To decompose each DWA into a micro-action sequence we applied a consensus multi-agent LLM pipeline. At an output scale of 15,817 actions, per-item expert rating is infeasible; quality control therefore rests on inter-model agreement at generation time, with human validation entering downstream through a 150-action stratified audit of the intelligence-type labels (see `Intelligence-type labelling'). The decomposition itself is quality-controlled by consensus only, and the resulting sequences are canonical task scripts as expressed in occupational language, not observations of performed work; the Discussion returns to this distinction. For every DWA, four locally deployed open-weight LLMs (Qwen, LLaMA, Gemma, Mistral families) independently produced candidate sequences of five to twelve steps against a fixed schema (final consensus sequences range from 5 to 13 steps); each model then read the full set of drafts and produced a synthesized version; finally each model voted on the best of the four synthesized candidates. Auto-resolution at 3:1 or stronger consensus accepted the majority draft directly ($n = 1{,}564$ DWAs). The 502 contested cases were arbitrated by three frontier-class models under the same voting scheme; 397 reached a 2-of-3 consensus and entered the dataset. The remaining 126 cases (105 with no majority agreement at either stage, plus 21 with a failed vote return) were dropped to preserve label quality, giving the conservation identity $1{,}564 + 397 + 126 = 2{,}087$ (model identifiers, decoding parameters, and prompt provenance in Supplementary Note~9). The discarded DWAs do not differ from the retained 1,961 in exposure (mean OAI $0.276$ vs $0.291$; Kolmogorov--Smirnov $D = 0.042$, $P = 0.98$), so the exclusion does not bias the polar contrasts.

The resulting dataset comprises 1,961 DWAs decomposed into 15,817 micro-actions. Each micro-action carries five structured fields: \texttt{step\_order} (position 1--13), \texttt{action\_description} (free text), \texttt{mapped\_stage} (one of five canonical process stages), \texttt{cognitive\_or\_physical} (Cognitive, Physical, Mixed, or Other, after normalisation of 39 LLM-emitted lexical variants), and \texttt{key\_challenge} (free text). The mean DWA contains 8.07 micro-actions (s.d.\ 1.34).

\subsection*{Embedding and dimensionality reduction}

Every action description was embedded into 768-dimensional vectors with the off-the-shelf \texttt{sentence-transformers/all-mpnet-base-v2} encoder \citep{reimers2019sentencebert, song2020mpnet}, without truncation or fine-tuning; whether off-the-shelf semantic similarity is adequate for occupational micro-actions is itself an empirical question the downstream validation answers. Dimensionality was reduced with UMAP \citep{mcinnes2018umap} in two parallel configurations: 2-d for visualisation (\texttt{n\_neighbors}=30, \texttt{min\_dist}=0.1, cosine metric) and 5-d for clustering (\texttt{min\_dist}=0.0), both at fixed \texttt{random\_state}=42. Five dimensions is a conservative compromise between preserving locally separable structure and keeping HDBSCAN's mutual-reachability distance well behaved; higher targets grew the cluster count without improving cluster TF-IDF interpretability. Both outputs were cached so that no UMAP stochasticity contaminates the statistical tests.

\subsection*{Two-stage clustering}

Stage one applied HDBSCAN \citep{campello2013hdbscan} to the 5-d space (\texttt{min\_cluster\_size}=100, \texttt{min\_samples}=20, EOM selection), identifying 35 micro-clusters containing 10,180 actions and leaving 5,637 actions (35.6\%) in the noise label. We deliberately did not force noise points into nearest clusters; the Results treat this group as a structural finding. Stage two built a 15-dimensional feature vector per micro-cluster (5-d UMAP centroid, five process-stage shares, four cognitive-physical shares, mean step order), z-scored each dimension, and computed Ward linkage \citep{murtagh2014ward} on the $35 \times 15$ matrix. Dendrogram inspection showed the largest interpretable jump structure at $K = 7$; at coarser cuts the eventual cognitive classes (M5--M7) or physical classes (M1, M2) remain conflated, while cuts beyond $K = 7$ produce singletons without new headline structure. The full trace, and the resolution sweep over $K = 7$--$15$ used in the robustness analysis, are given in Supplementary Note~1. A DWA whose largest class share falls below 0.40 is assigned to a separate mixed-DWA cell (96 DWAs, 4.9\%); micro-action sequences are constitutively spread across stages, so a stricter majority rule would over-classify well-typed DWAs as mixed.

\subsection*{Exposure indicators}

Our first exposure indicator is a DWA-level Automation Index (OAI) constructed by a tech-risk dual-factor procedure in our own companion preprint \citep{gao2026boundedrisk}: a four-LLM ensemble assessment of technical capability $T_i \in \{0,\dots,3\}$ is combined with a business-risk score $R_i \in \{1,\dots,5\}$ through a bottleneck (Leontief-style) mapping in which high risk vetoes automation regardless of capability, producing values on the discrete set $\{0, 0.3, 0.5, 0.7, 1.0\}$. In its original construction the per-DWA scores were calibrated against a 31-expert variance-based stratified human-in-the-loop panel on a 100-DWA sample. The OAI covers all 2,087 baseline DWAs; the 1,961 clustered DWAs match with zero missing values.

The projection is external validation, not an input: no OAI value, DWA-level score, or occupation-level score enters the clustering pipeline at any stage, so the typology cannot be circular with respect to the exposure indicators projected onto it.

Our second, co-primary indicator is the per-task GPT-4 exposure rating of \citet{eloundou2023gpts}, mapped from five tiers to $\{0, 0.25, 0.5, 0.75, 1.0\}$ and joined to DWAs through the O*NET 30.2 task--DWA crosswalk by equal-weight averaging (coverage 1,958 of 1,961 DWAs). For the temporal analysis we additionally aligned Felten's AIOE \citep{felten2021occupational} (SOC-level, replicated across constituent DWAs; coverage 97.6\%) and Frey--Osborne's computerisation probability, parsed directly from the original Oxford Martin appendix \citep{frey2013oxfordmartin} and verified against the compilation embedded in \citet{eloundou2023gpts} at Spearman $\rho = 1.000$ across 653 matched SOC codes. For the occupation-level comparison, O*NET-SOC codes were truncated to six-digit SOC and matched to the 702-row Frey--Osborne table (618 codes matched); occupation-level OAI is the task-coverage-weighted index of the companion preprint, an occupation's Linguistic composition is its $\omega_L$ share from the released occupation table, and values were averaged where several O*NET-SOC codes share one six-digit SOC.

\subsection*{Statistical tests}

The primary battery comprised (1) Kruskal--Wallis non-parametric ANOVA across the eight analysis groups; (2) pairwise Mann--Whitney $U$ with Bonferroni correction over all 28 pairs; (3) effect sizes as Cliff's $\delta$ and Cohen's $d$; (4) Hartigan's dip test of unimodality \citep{hartigan1985dip} (heuristic on the discrete OAI support; Supplementary Note~3); and (5) two-sample Kolmogorov--Smirnov tests between the substrate and the population. The robustness battery comprised (6) a resolution sweep re-cutting the dendrogram at $K = 7, 8, 10, 12, 15$ and recomputing all polar statistics; (7) two-one-sided equivalence tests (TOST) on the 15 middle pairs at margin $d = 0.2$ ($\delta = 0.060$ on the OAI scale) and at an absolute margin of $\pm 0.05$, with the extreme pair as a manipulation check (TOST must and does reject equivalence there); and (8) reproduction of the full test battery under the three external indicators. As (9), a permutation test drew 20,000 random partitions of the DWA-level exposure values into eight groups of the observed sizes and benchmarked the observed polar gap, middle-band compression, and the bimodality coefficient of the eight group means against this null. All random-state-sensitive procedures used \texttt{random\_state}=42.

\subsection*{Intelligence-type labelling}

Each micro-action was classified into one of four intelligence types (Linguistic, Multimodal Perception, Embodied, or Human-Bound) by prototype-anchored nearest-centroid assignment: eight unambiguous prototype actions per class were embedded, class centroids computed, and each action assigned to its nearest centroid by cosine similarity. The four types operationalise the four primary technical bottlenecks of contemporary AI; they are an explanatory lens, not a proposed taxonomy, and alternative decompositions \citep{morris2023levels, oecd2025capability, openai2024fivelevels} would be legitimate extensions (Supplementary Note~6).

For this supervised step we used the \texttt{BAAI/bge-large-en-v1.5} encoder \citep{xiao2023bge} rather than the MPNet encoder used for unsupervised clustering, on the basis of a 150-action stratified human audit: the authors hand-labelled a stratified sample of 150 actions against the operational class definitions, and BGE labels matched these author labels at $\kappa = 0.893$ (92.0\% accuracy) against MPNet's $\kappa = 0.769$ (82.7\%), with the gap concentrated on the 30 audit rows where the encoders disagreed. Because BGE was both selected and evaluated on these same 150 rows, its $\kappa$ is an optimistic estimate; the selection-independent anchors are MPNet's $\kappa = 0.769$ and the 97.5\% joint accuracy on the 120 rows where the encoders agree. MPNet results are retained as a robustness check and match qualitatively (Supplementary Note~5). To move the human anchor off the authors, three annotators unaware of the model labels and study hypotheses independently re-annotated a fresh stratified 150-action sample and rated the decomposition fidelity of 79 DWAs. The four-category construct is highly reproducible across these independent raters (human--human Fleiss $\kappa = 0.90$; pairwise Cohen $\kappa$ 0.86--0.98; 89\% unanimous), confirming that its reliability is not an artefact of author involvement. The automatic BGE labels, however, match the independent annotators at $\kappa = 0.63$ (mean over three raters; 74\% agreement with the human-majority label), substantially below the in-loop author audit: relative to human consensus the encoder over-assigns Multimodal Perception and Human-Bound and under-assigns Linguistic. We therefore treat per-action intelligence types as reliable at the class and aggregate level but noisy individually; the class-level contrasts that carry the mechanistic argument (Supplementary Fig.~S5) do not depend on any single action's label. Decomposition fidelity is high (mean completeness $4.98$, faithfulness $4.63$, granularity $4.78$ on a 1--5 scale; no item scored below $3$ by any rater), but annotators consistently flagged two systematic failure modes. The first is scope drift beyond the DWA verb, where ``plan'', ``direct'', or ``assess'' activities are decomposed into downstream execution. The second is occasional ceremonial or navigational filler steps. We document both as targets for a future dataset revision (Supplementary Note~5).

\subsection*{Data availability}

O*NET 30.2 is publicly available from the US Department of Labor. The Eloundou et al.\ task ratings, Felten et al.\ AIOE scores, and the Oxford Martin Frey--Osborne appendix are available from their original publications. The Final Golden Dataset of 15,817 micro-actions, all cluster assignments, the micro-cluster-to-macro-class mapping, all projected indicator values, and an occupation-level table of intelligence-type compositions ($\bm{\omega}$ shares and substrate share for all 923 occupations) are openly available as The Micro-Action Dataset on Zenodo at \url{https://doi.org/10.5281/zenodo.21395793} (v1.0.0; all versions: \url{https://doi.org/10.5281/zenodo.21395792}).

\subsection*{Code availability}

The full pipeline (decomposition, embedding, clustering, projection, and statistical tests) is available at \url{https://github.com/ShuyaoGao/occupational-micro-actions}.

\bibliography{references}

\end{document}